\documentclass[aps,superscriptaddress,nofootinbib]{revtex4}
\usepackage{graphicx}	
\usepackage{dcolumn}	
\usepackage{bm}		
\usepackage{amsmath}
\usepackage{amsfonts}
\usepackage{amssymb}
\usepackage{hyperref}
\usepackage{mathtools}
\usepackage{setspace}	
\usepackage{color}
\usepackage{dsfont}
\usepackage{pdfpages}

\definecolor{green}{rgb}{0,.45,0}
\definecolor{orange}{rgb}{1,0.5,0}

\newcommand{\be}{\begin{equation}}
\newcommand{\ee}{\end{equation}}
\newcommand{\ba}{\begin{eqnarray}}
\newcommand{\ea}{\end{eqnarray}}
\newcommand{\la}{\langle}
\newcommand{\ra}{\rangle}
\newcommand{\di}{\mathrm{d}}

\begin{document}
\title{\boldmath 
	Determination of $J/\psi$
	chromoelectric polarizability from lattice data}

\author{Maxim V.~Polyakov}
	\affiliation{Petersburg Nuclear Physics Institute, 
		Gatchina, 188300, St.~Petersburg, Russia}
	\affiliation{Institut f\"ur Theoretische Physik II, 
		Ruhr-Universit\"at Bochum, D-44780 Bochum, Germany}
\author{Peter~Schweitzer}
	\affiliation{Department of Physics, University of Connecticut, 
		Storrs, CT 06269, USA}

\date{July 2018}  

%

\begin{abstract} 
The chromoelectric polarizability of $J/\psi$ is extracted 
from lattice QCD data on the nucleon-$J/\psi$ potential in 
the heavy quark limit. The value of 
$\alpha(1S) = (1.6 \pm 0.8) \,{\rm GeV}^{-3} $ is obtained.
This value may have a systematic uncertainty due to 
lattice artifacts which cannot be estimated at present, but 
will become controllable in future studies.
We also comment on the possibility of hadrocharmonia.
\end{abstract}

\maketitle


\section{Introduction}
\label{Sec-1:Introduction}

The chromoelectric polarizability $\alpha$ of a hadron describes the 
hadron's effective interaction with soft gluonic fields. This property 
is analogous to the electric polarizability quantifying the response of 
a neutral atom placed in an external electric field, which describes 
the emergence of induced dipole moments and van der Waals forces.

The chromoelectric polarizabilities of charmonia are important quantities
in the heavy quark effective theory. Among their most interesting applications
are studies of hadrocharmonia: when the compact charmonium penetrates a light 
hadron, its interaction with the soft gluon fields inside the hadron 
is systematically described in terms of a multipole expansion 
\cite{Gottfried:1977gp,Voloshin:1978hc}. 
The strength of the effective interaction between a charmonium and 
a light hadron is determined by the chromoelectric polarizability
of the charmonium \cite{Voloshin:1979uv,Voloshin:2007dx}.
If this effective interaction is strong enough, bound states emerge: 
hadrocharmonia 
\cite{Voloshin:2007dx,Dubynskiy:2008mq,Sibirtsev:2005ex,Eides:2015dtr}.
The binding of $J/\psi$ in nuclear medium and nuclei was also studied 
\cite{Brodsky:1989jd,Luke:1992tm,Kaidalov:1992hd,Tsushima:2011kh}. 

Other important applications of chromoelectric polarizabilities include
the description of hadronic transitions between charmonium resonances
\cite{Voloshin:1980zf,Novikov:1980fa} and the interaction of slow 
charmonia with a nuclear medium. The chromoelectric polarizabilities
also play a vital role for the understanding of photo-production and 
hadro-production of charmonia and charmed hadrons on nuclear targets
with important applications for the diagnostics of the creation of 
quark gluon plasma in heavy-ion collisions, see \cite{Matsui:1986dk,
Barnes:2003vt,Brambilla:2004wf,Shuryak:2014zxa,Braun-Munzinger:2015hba}
and references therein.

Despite their importance little is known about the phenomenological values
of these nonperturbative charmonium properties. Only on the transitional 
chromoelectric polarizability $\alpha(2S\to1S)$ is some information available 
\cite{Voloshin:2007dx}. 
The value of $\alpha(1S)$ could in principle be inferred from the rare 
decay $J/\psi\to\pi\pi\ell^+\ell^-$ with soft pions \cite{Voloshin:2004un},
but such an analysis is challenging and  $\alpha(1S)$ is not yet known.

A nonperturbative determination of $\alpha(1S)$ is therefore of great 
importance. In this work we present a method to determine $\alpha(1S)$ 
from lattice QCD calculations of the $J/\psi$-nucleon potential. We
estimate conservatively the theoretical uncertainties which are
associated with underlying assumptions, and discuss critically 
how these assumptions can be tested with future lattice data.
We also comment on the possibility of nucleon-$\psi(2S)$ 
bound states.

\section{The effective quarkonium-baryon interaction}
\label{Sec-2:Veff}

The interaction of a heavy quarkonium with a baryon is dominated 
in the heavy quark limit by the emission of two virtual color-singlet 
chromoelectric dipole gluons \cite{Voloshin:1979uv,Voloshin:2007dx} 
and described, for $S$-wave quarkonia, by an effective potential in terms 
of the quarkonium chromoelectric polarizability $\alpha$ and energy-momentum 
tensor (EMT) densities of the baryon as \cite{Eides:2015dtr}
\be\label{Eq:Veff}
	V_{\rm eff}(r) = -\,\alpha\;\frac{4\pi^2}{b}\,
	\frac{g_c^2}{g_s^2}\,\biggl(\nu\,T_{00}(r)-3\,p(r)\biggr) 
	\, , \quad \quad
	\nu = 1+\xi_s\,\frac{b\,g_s^2}{8\pi^2} \; ,
\ee
where $b = (\frac{11}{3} N_c-\frac{2}{3}\,N_f)$ is the leading coefficient 
of the Gell-Mann--Low function, $g_c$ ($g_s$) is the strong coupling 
constant renormalized at the scale $\mu_c$ ($\mu_s$) associated with 
the heavy quarkonium (baryon) state. The parameter $\xi_s$ denotes 
the fraction of the baryon energy carried by gluons at the scale 
$\mu_s$ \cite{Novikov:1980fa}. In Eq.~(\ref{Eq:Veff})
$T_{00}(r)$ and $p(r)$ are the energy density and pressure inside
the baryon \cite{Polyakov:2002yz}, which satisfy respectively 
\be\label{Eq:mass-von-Laue}
	\int\di^3r\:T_{00}(r)=M_B 	\, , \quad \quad
	\int \di^3 r\:p(r) = 0 	\, ,
\ee
where $M_B$ denotes the mass of the baryon.
The derivation of Eq.~(\ref{Eq:Veff}) is justified in the limit that
the ratio of the quarkonium size is small compared to the effective gluon 
wavelength \cite{Voloshin:2007dx}, and a numerically small term proportional 
to the current masses of the light quarks is neglected.

Due to Eq.~(\ref{Eq:mass-von-Laue}) the effective potential has the
following normalization and mean square radius
\be\label{Eq:Veff-norm}
	\int\di^3r\;V_{\rm eff}(r) = -\,\alpha\;
	\frac{4\pi^2}{b}\,\frac{g_c^2}{g_s^2}\;\nu\,M_B \, , \quad \quad
	\la r^2_{\rm eff}\ra \equiv 
	\frac{\int\di^3r\;r^2V_{\rm eff}(r)}{\int\di^3r\;V_{\rm eff}(r)}
	= \la r_E^2\ra - \frac{12\, d_1}{5\nu M_B^2}
\ee
with the mean square radius of the energy density
$\la r_E^2\ra = \int\di^3r\;r^2T_{00}(r) /M_B$ and the 
$D$-term $d_1=\frac54\,M_B\int\di^3 r\:r^2p(r)$ 
\cite{Polyakov:2002yz,Polyakov:1999gs}.
Using the normalization condition for $V_{\rm eff}$ in
Eq.~(\ref{Eq:Veff-norm}) to eliminate the ratio $(g_c/g_s)^2$
from Eq.~(\ref{Eq:Veff}) and exploring the large-$r$ behavior of 
$T_{00}(r)$ and $p(r)$ derived in \cite{Goeke:2007fp} we obtain 
the following expression for the long-distance behavior of 
$V_{\rm eff}(r)$ in the chiral limit, which is convenient for
our purposes:
\be\label{Eq:Veff-large-r}
	V_{\rm eff}(r) = \frac{27}{16\,\pi^2}\;\frac{1+\nu}{\nu}\;
	\frac{g_A^2}{M_B F_\pi^2}\;\frac{1}{r^6}\;
	\int\di^3r^\prime\,V_{\rm eff}(r^\prime) \quad \mbox{for $r$ large,}
\ee
where $F_\pi=93\,{\rm MeV}$ is the pion decay constant, and $g_A$ is the 
axial coupling constant with $g_A=1.26$ for the nucleon. Notice that this
result refers to the leading order of the expansion in a large number of 
colors $N_c$ \cite{Goeke:2007fp} with $N_c\to\infty$ taken first, and 
$m_\pi\to 0$ taken second (in general these limits do not commute).
For finite $m_\pi$ the behavior is 
$V_{\rm eff}(r) \propto \exp(-2m_\pi r)/r^2$ at $r\gg 1/m_\pi$ 
\cite{Goeke:2007fp}.

\section{Chromoelectric polarizabilities}
\label{Sec-2b:polarizabilities}

The chromoelectric polarizabilities $\alpha$ are important properties 
of quarkonia. Little is known about them especially for charmonia, 
except that the chromoelectric polarizabilities of 
$J/\psi$ and $\psi^\prime$, $\alpha(1S)$ and $\alpha(2S)$, are real 
and positive, and satisfy the Schwarz inequality 
$\alpha(1S)\,\alpha(2S) \ge \alpha(2S\to1S)^2$
\cite{Voloshin:2007dx}.
The chromoelectric polarizabilities were calculated in the large-$N_c$ 
limit in the heavy quark approximation \cite{Peskin:1979va}.
Applying the results to the charmonium case yields \cite{Eides:2015dtr}
\begin{subequations}
\label{Eq:alpha-guideline}
\begin{align}
	\label{Eq:alpha-1S}
	\alpha(1S)_{\rm pert.} \approx & \;\;\;\;\,0.2\,{\rm GeV}^{-3} \;, \\
	\label{Eq:alpha-2S}
	\alpha(2S)_{\rm pert.} \approx & \;\;\;\;\;\,12\, {\rm GeV}^{-3} \;, \\
	\label{Eq:alpha-2S-to-1S}
	\alpha(2S\to1S)_{\rm pert.}\approx & -0.6\, {\rm GeV}^{-3}\;.
\end{align}
\end{subequations}
Independent phenomenological information on the value of the 
$2S \to 1S$ transition polarizability is available from analyses of 
data on the decay $\psi^\prime\to J/\psi \, \pi\pi$ \cite{Voloshin:2007dx}
\be\label{Eq:alpha-2S-to-1S-pheno}
	|\alpha(2S\to 1S)| \approx 2\,{\rm GeV}^{-3}\;
	\mbox{(phenomenology)}.
\ee
In the heavier bottomonium system $1/N_c$ corrections to $\alpha(1S)$ 
are of ${\cal O}(5\,\%)$ \cite{Brambilla:2015rqa}. In the charmonium 
system presently no information is available on the chromoelectric 
polarizabilities besides the perturbative estimates \cite{Peskin:1979va} 
and the phenomenological value for the $2S\to1S$ polarizability 
\cite{Voloshin:2007dx} which is 
only in rough agreement with the perturbative prediction, see
Eq.~(\ref{Eq:alpha-2S-to-1S}) vs (\ref{Eq:alpha-2S-to-1S-pheno})
[notice that $\pi\pi$ final state interactions \cite{Guo:2006ya} 
may reduce the value in Eq.~(\ref{Eq:alpha-2S-to-1S-pheno})].

In this situation, independent information on the chromoelectric 
polarizabilities of charmonia is of importance. 

\section{Extraction of the chromoelectric polarizability of 
\boldmath $J/\psi$}
\label{Sec-4:alpha}

The recent lattice QCD data on the effective charmonium-nucleon 
interaction \cite{Sugiura:2017vks} put us in the position to extract 
the chromoelectric polarizability $\alpha(1S)$ of $J/\psi$.
This nonperturbative determination of $\alpha(1S)$ 
warrants a study, even though the lattice data \cite{Sugiura:2017vks} 
(published in a conference proceeding) may have unestimated 
systematic uncertainties. 
The results of Ref.~\cite{Sugiura:2017vks} were obtained using $2 + 1$ 
flavor full QCD gauge configurations which were simulated with a Wilson 
clover quark action on a $16^3\times32$ lattice with lattice spacing 
$a = 0.1209\,{\rm fm}$.
Using this action for heavy quarks ``may bring large discretization errors'' 
as stressed in \cite{Sugiura:2017vks}. Another concern are the unphysical
light quark masses used in \cite{Sugiura:2017vks} which correspond to a
pion mass of $m_\pi=875\,{\rm MeV}$. The results of \cite{Sugiura:2017vks}
are in qualitative agreement with earlier studies in quenched lattice QCD
\cite{Kawanai:2010ev}. Until future lattice QCD studies performed
with physical light quark masses on finer lattices or with relativistic 
heavy quark action for charm, we have to keep these points in mind
as unestimated potential systematic uncertainties in our extraction.

The extraction assumes that the charm-quark mass is sufficiently large 
to neglect heavy quark mass corrections, which can be tested with future 
lattice QCD data. Although below we will see that the lattice data 
are compatible with this assumption, presently also this point has to
be kept in mind as a potential uncontrolled systematic uncertainty.

From Eq.~(\ref{Eq:Veff-norm}) we obtain (here $M_N$ denotes the nucleon mass)
\be\label{Eq:alpha}
	\alpha = -\;\frac{b}{4\pi^2\,\nu\,M_N}\;
	\frac{g_s^2}{g_c^2} \int\di^3r\;V_{\rm eff}(r) \,. 
\ee
Let us discuss the different factors which play a role in the
extraction of $\alpha$ and their uncertainties.

The coefficient $\nu$ introduced in Eq.~(\ref{Eq:Veff}) was
estimated on the basis of the instanton liquid model of the QCD 
vacuum and the chiral quark soliton model, where the strong 
coupling constant freezes at a scale set by the nucleon size at 
$g_s^2/(4\pi) \approx 0.5$. Assuming $\xi_s\approx 0.5$ as 
suggested by the fraction of nucleon momentum carried by gluons 
in DIS at scales comparable to $\mu_s$ one obtains the value
$\nu \approx 1.5$ \cite{Eides:2015dtr}. 
This is supported by the analysis of the nucleon mass decomposition 
\cite{Ji:1994av} with $\xi_s\approx \frac13$ leading to $\nu \approx 1.4$.
Based on these results we will use
\be\label{Eq:nu}
	\nu \approx 1.5\pm0.1
\ee 
in this work. Let us remark that a similar result $\nu =(1.45\dots1.6)$
was obtained for the pion in Ref.~\cite{Novikov:1980fa}. 

In order to estimate the factor $g_s^2/g_c^2$ we use two extreme approaches. 
One estimate is based on effective nonperturbative methods. 
For that we use the nonperturbative result $g_s^2/(4\pi)\approx0.5$ from 
the instanton vacuum model mentioned above which refers to a low scale of 
the nucleon, see above. Interestingly, phenomenological calculations of 
charmonium properties require $g_c^2/(4\pi) = 0.5461$ at a scale 
associated with charmonia \cite{Barnes:2005pb}. This indicates that 
$g_s^2/g_c^2\sim 1$ is a reasonable assumption \cite{Eides:2015dtr}. 
Another ``extreme'' result is provided by the leading-order QCD 
running coupling constant. We follow Ref.~\cite{Gluck:1998xa} 
where the description of the strong coupling constant was optimized 
to guarantee perturbative stability down to a low initial scale 
$\mu_{\rm LO}^2=0.26\,{\rm GeV}^2$ of the parametrizations for the 
unpolarized parton distribution functions. 
In this way we obtain $g_s^2/(4\pi)=0.46$ at a scale set by the 
nucleon mass, while $g_c^2/(4\pi)=(0.27\dots0.36)$ depending on whether
one evaluates the running coupling constant at the scale $m_c$ or $2m_c$ 
(the leading-order derivation of Eq.~(\ref{Eq:Veff}) does not fix the
scale, and both choices are equally acceptable). In this way we obtain the 
``leading-order perturbative estimate'' $g_s^2/g_c^2\sim (1.3\dots1.7)$.
This indicates that this quantity is associated with a substantial 
theoretical uncertainty. In order to cover both extreme cases, 
we will assume that
\be\label{Eq:ratio-gs-gc}
	\frac{g_s^2}{g_c^2} \approx 1.37\pm 0.37 \;.
\ee 

The information on $\int\di^3r\,V_{\rm eff}(r)$ is obtained from
the lattice QCD calculation \cite{Sugiura:2017vks} performed with 
unphysical light quark masses such that $m_\pi=875\,{\rm MeV}$ and 
$M_N=1816\,{\rm MeV}$ but with a physical value of $m_c$. In 
the heavy quark limit the effective potential factorizes in the 
chromoelectric polarizability $\alpha$ and nucleonic properties,
and we may expect the extracted value of $\alpha$ to be 
weakly affected by the unphysical light quark masses. 
(The heavy quark mass corrections might be sensitive to light quark masses.
This is part of the currently uncontrolled systematic uncertainties, 
which can be revisited in future when lattice calculations with physical 
light quark masses will become available for $V_{\rm eff}$.)

\begin{figure}[b!]

\vspace{5mm}

\centering
\includegraphics[height=5.25cm]{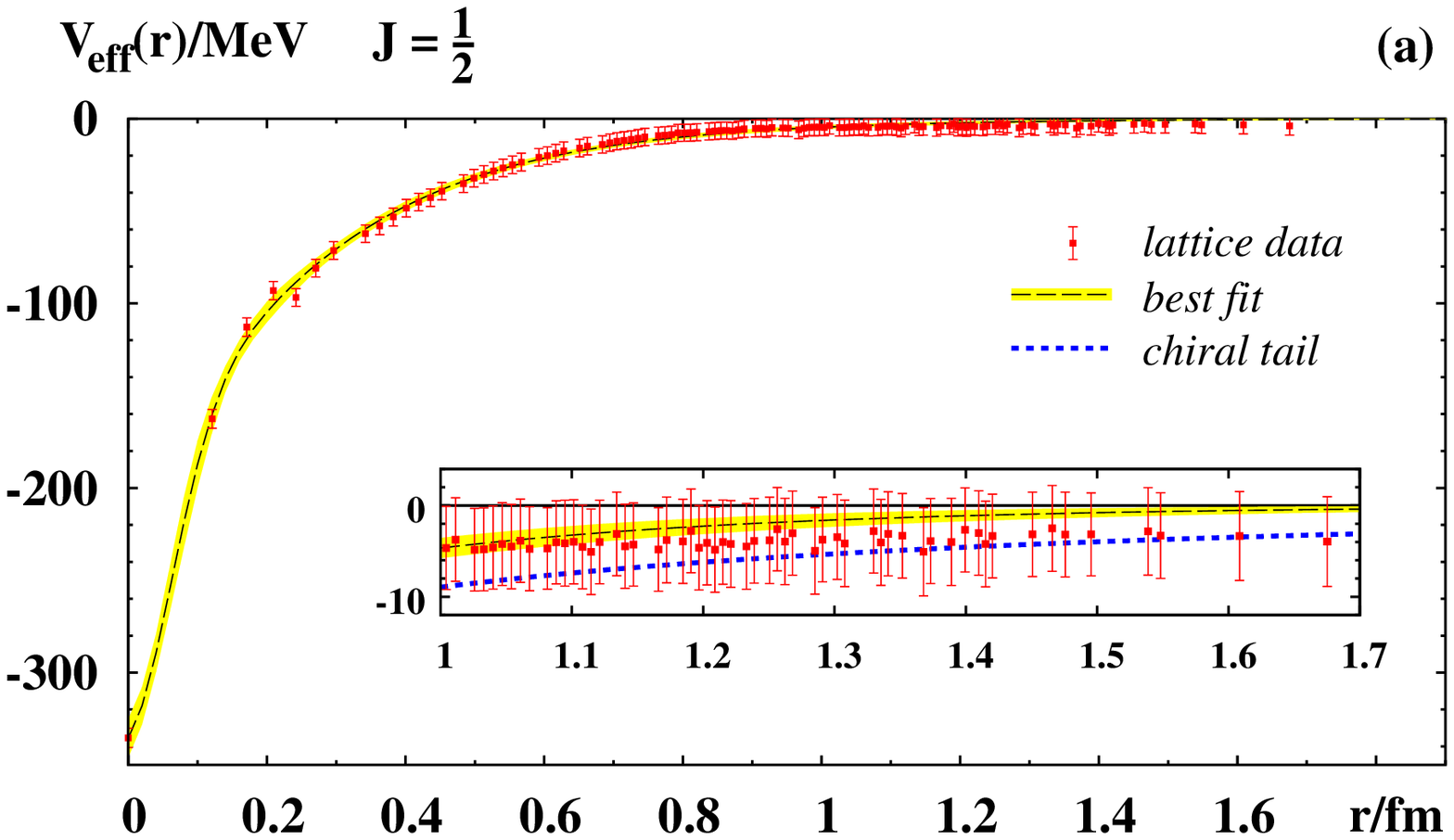}%
\includegraphics[height=5.25cm]{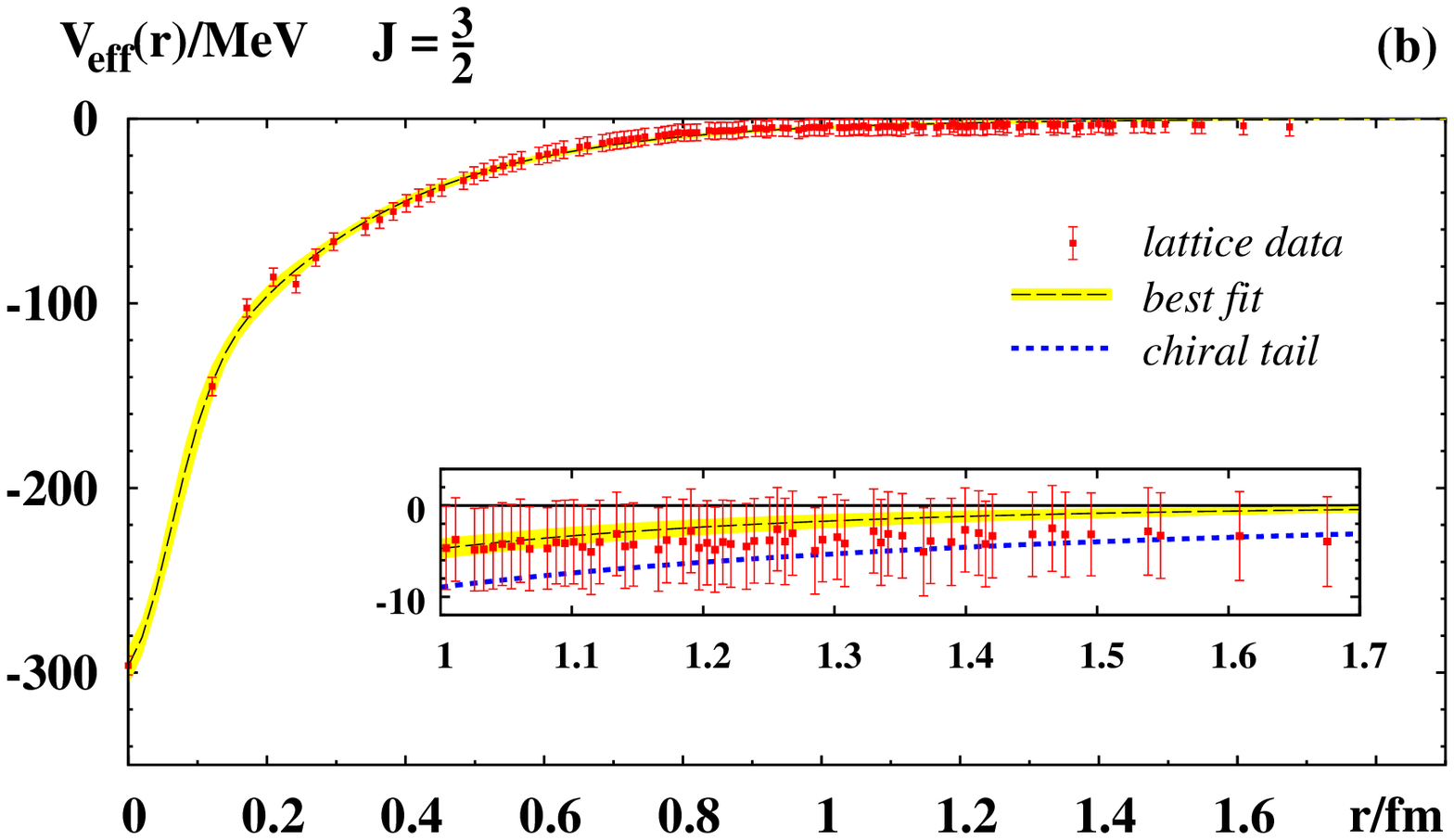} 

\vspace{5mm}

\caption{\label{FIG-01:Veff-lattice}
	Effective $J/\psi$-nucleon potential $V_{\rm eff}(r)$ 
	as function of $r$ from the lattice QCD calculation 
	\cite{Sugiura:2017vks} and the best fits 
	(\ref{Eq:fit-form}-\ref{Eq:fit-result-32})
	in the channels: (a) $J=\frac12$, and (b) $J=\frac32$.
	The shaded areas show the 1-$\sigma$ regions of the fits. The 
	insets show the regions of $1\,{\rm fm} < r < 1.7\,{\rm fm}$
	where the available lattice data are compatible 
	within error bars also with zero or with chiral predictions.}
\end{figure}

In the lattice calculation $V_{\rm eff}(r)$ was computed in the region
$0\le r\le 1.7\,{\rm fm}$ in the angular momentum channels $J=\frac12$
and $J=\frac32$ as shown in Fig.~\ref{FIG-01:Veff-lattice}.
The lattice data in both channels can be fitted with functions of the form 
\be\label{Eq:fit-form}
	V_{\rm eff}(r) 	= 	     C_0\, e^{-\,\frac{r^{ }}{r_0}}\;
	\frac{1}{1+\frac{r^2}{r_1^2}}+C_2\, e^{-\,\frac{r^2}{r_2^2}}\,,
\ee
where the first term is defined such that at large $r$ it has the form 
dictated by chiral symmetry \cite{Goeke:2007fp}, while the second term 
constrains the parametrization in the small-$r$ region.
(We are not aware of a deep physical reason why the second
term should be Gaussian, besides the fact that among the Ans\"atze 
we explored it yields the lowest $\chi^2$, see below.)
The best fit parameters in the channel $J=\frac12$ are as follows:
\begin{align}
    C_0^{(1/2)} = {-(178.2\pm 3.7)\,{\rm MeV}}  \, ,	
 && r_0^{(1/2)} = { (0.573\pm 0.065)\,{\rm fm}} \, ,	
 && r_1^{(1/2)} = { (0.429\pm 0.041)\,{\rm fm}} \, ,	
	\nonumber\\
    C_2^{(1/2)} = {-(157.4\pm 4.3)\,{\rm MeV}} \, ,	
 && r_2^{(1/2)} = { (0.091\pm 0.003)\,{\rm fm}} \, ,	
 && \chi^2_{\rm d.o.f.} = {0.18} \, ,			
	\label{Eq:fit-result-12}
\end{align}
The best fit parameters in the channel $J=\frac32$ are as follows:
\begin{align} 
    C_0^{(3/2)} = {-(160.4\pm 3.3)}\,{\rm MeV} \, , 	
 && r_0^{(3/2)} = { (0.619\pm 0.073)}\,{\rm fm}\, , 	
 && r_1^{(3/2)} = { (0.426\pm 0.039)}\,{\rm fm}\, ,  
	\nonumber\\
    C_2^{(3/2)} = {-(136.0\pm 3.9)}\,{\rm MeV}\, , 	
 && r_2^{(3/2)} = { (0.088\pm 0.004)}\,{\rm fm}\, ,	
 && \chi^2_{\rm d.o.f.} = {0.17}  		 \, .   
	\label{Eq:fit-result-32}
\end{align}
The fits are shown in Fig.~\ref{FIG-01:Veff-lattice}.
Several remarks are in order.

First, 
the potentials in both channels are very similar, and agree
with each other within $\pm5\,\%$ relative accuracy. In fact,
except for the point at $r=0$ both lattice data sets are 
compatible with each other within error bars. Let us remark that, 
if heavy quark mass corrections play a role, one should expect
them to have an impact especially in the region of small 
$r\lesssim 1/m_c \approx 0.13\,{\rm fm}$.
The independence of $V_{\rm eff}(r)$ of $J=\frac12$ or $\frac32$ 
is an important consistency check of our approach.
The effective potential is universal in our approach, and differences
due to different $J$ are expected to be suppressed in the heavy quark
limit, as we observe. Thus, we have no indication that heavy quark mass 
corrections are significant for $V_{\rm eff}(r)$ in the charmonium system. 
As mentioned above, this point can be tested quantitatively with future 
lattice data.

Second, chiral symmetry dictates $r_0=(2m_\pi)^{-1}=0.11\,{\rm fm}$.
The fits are a factor of 5 off. Notice, however, that the lattice data 
clearly constrain $V_{\rm eff}(r)$ in both channels only up to about 
$r\lesssim 1\,{\rm fm}$. It is likely that this limited $r$-region
does not extend far enough to see the chiral asymptotics. Indeed,
for $1\,{\rm fm} < r < 1.7\,{\rm fm}$ the lattice data on $V_{\rm eff}(r)$ 
are actually compatible with zero within error bars, see the insets in
Fig.~\ref{FIG-01:Veff-lattice}. Notice, however, that a fit with
the fixed parameter $r_0=(2m_\pi)^{-1}$ 
(with $m_\pi=875\,{\rm MeV}$ here) has still an excellent 
$\chi^2$ per degree of freedom of $\chi^2_{\rm d.o.f.}= 0.4$
for both channels.
This is remarkable and indicates that the lattice data are compatible
with chiral symmetry.

Third, we explored also other shapes for the fit functions with
practically no difference in the region $r\lesssim 1\,{\rm fm}$
where the lattice data have the strongest constraining power.
We will comment below on the region $r>1\,{\rm fm}$.

In order to evaluate $\int\di^3r\,V_{\rm eff}(r)$ we consider separately 
the region $r < 1\,{\rm fm}$ where the lattice data are clearly non-zero, 
and $r \ge 1\,{\rm fm}$ where the lattice data are compatible 
with zero within error bars (including the region $r>1.7\,{\rm fm}$ with
no available lattice data), see the inset in 
Fig.~\ref{FIG-01:Veff-lattice}. In the region $r < 1\,{\rm fm}$
the fits in Eqs.~(\ref{Eq:fit-form}--\ref{Eq:fit-result-32})
yield
\be\label{Eq:integral-small-r}
	\int\limits_{\!\!\!r<1\,\rm fm\!\!\!\!\!\!}\di^3r\,V_{\rm eff}(r) = 
	\begin{cases}
	(-9.3\pm 0.8)\,{\rm GeV}^{-2} & \mbox{for} \quad J=\frac12\,,\\
	(-8.9\pm 0.8)\,{\rm GeV}^{-2} & \mbox{for} \quad J=\frac32\,.
	\end{cases}
\ee
The uncertainty of these results is due to the statistical 
uncertainty of the lattice data. We tried several other fit
Ans\"atze which all had larger $\chi^2_{\rm d.o.f.}$, and
gave results compatible with (\ref{Eq:integral-small-r}) 
within statistical error bars. The systematic uncertainty
due to the choice of fit Ansatz is therefore negligible
compared to the statistical uncertainty of the fits.

In the region $r>1\,{\rm fm}$ systematic uncertainties due to the
choice of fit Ansatz are not negligible. The form (\ref{Eq:fit-form})
of the best fit is well motivated by chiral symmetry. But the lattice 
data \cite{Sugiura:2017vks} have a modest constraining power for 
$1\,{\rm fm} < r < 1.7\,{\rm fm}$, and no lattice data are 
available beyond that. To proceed we assume that the 
fits~(\ref{Eq:fit-form}--\ref{Eq:fit-result-32})
give useful estimates for the central values of contributions 
from $r>1\,{\rm fm}$ to the integrals over $V_{\rm eff}(r)$, 
and assign a systematic error by using two extreme estimates.
For the first estimate we approximate $V_{\rm eff}(r) = 0$ for 
$r\ge 1\,{\rm fm}$, which fits the lattice data in the region 
$1\,{\rm fm} < r < 1.7\,{\rm fm}$ with a $\chi^2_{\rm d.o.f.}=0.7$, 
and certainly leads to overestimates of the contributions from the
large-$r$ region to $\int\di^3r\,V_{\rm eff}(r)$ in both channels. 
For the second extreme estimate we
assume $V_{\rm eff}(r)\propto 1/r^6$ with the coefficient given
by Eq.~(\ref{Eq:Veff-large-r}). Notice that the coefficient strictly 
speaking needs the full result for $\int\di^3r\,V_{\rm eff}(r)$ which
we do not yet know. At this point one could design an iterative
procedure, but for our purposes it is sufficient to assume that
$\int\di^3r\,V_{\rm eff}(r)\approx -(10\dots20)\,{\rm GeV}^{-2}$. 
This is also compatible with the lattice data (a fit assuming 
$\int\di^3r\,V_{\rm eff}(r) = -15\,{\rm GeV}^{-2}$ has 
$\chi^2_{\rm d.o.f.}=0.20$ and is shown in Fig.~\ref{FIG-01:Veff-lattice}) 
and certainly leads to an underestimate of the large-$r$ contribution 
to the integral. To summarize, in the large-$r$ region we obtain
\be\label{Eq:integral-large-r-prepare}
	\int\limits_{\!\!\!r\ge 1\,\rm fm\!\!\!\!\!\!}\di^3r\,V_{\rm eff}(r) 
	=
	\begin{cases}
	0				&
	\mbox{$J=\frac12$, $\frac32\,$
	extreme estimate (i): $V_{\rm eff}(r)=0$ for $r>1\,{\rm fm}$,}\\
	{-(4.9\pm 3.4)}\,{\rm GeV}^{-2} 	& 
	\mbox{$J=\frac12$ $\to$
	extrapolation based on the best fit in 
	Eqs.~(\ref{Eq:fit-form},~\ref{Eq:fit-result-12}),}\\
 	{-(5.4\pm 3.9)}\,{\rm GeV}^{-2} 	& 
	\mbox{$J=\frac32$ $\to$
	extrapolation based on the best fit in 
	Eqs.~(\ref{Eq:fit-form},~\ref{Eq:fit-result-32}),}\\
	-(3.3\dots6.6)\,{\rm GeV}^{-2}	&
	\mbox{$J=\frac12$, $\frac32\,$
	extreme estimate (ii): $V_{\rm eff}(r)$ with ``chiral tail'' 
	for $r>1\,{\rm fm}$.}
	\end{cases}
\ee
We use the best fit results as central values and the extreme
estimates to assign a systematic uncertainty as follows
\be\label{Eq:integral-large-r}
	\int\limits_{\!\!\!r\ge 1\,\rm fm\!\!\!\!\!\!}\di^3r\,V_{\rm eff}(r) 
	= 	
	\begin{cases}
	 {-4.9\pm 3.4^{+4.9}_{-1.7}}\,{\rm GeV}^{-2} & \mbox{$J=\frac12$}\,, \\
	 {-5.4\pm 3.9^{+5.4}_{-1.2}}\,{\rm GeV}^{-2} & \mbox{$J=\frac32$}\,.
	\end{cases}	
\ee
Combining Eqs.~(\ref{Eq:integral-small-r},~\ref{Eq:integral-large-r})
the final result for the full integral of the effective potential is
\be\label{Eq:integral}
	\int\di^3r\,V_{\rm eff}(r)
	=
	\begin{cases}
  \,{(-14.2\pm 0.8^{+6.0}_{-3.8})}\,{\rm GeV}^{-2} & \mbox{$J=\frac12$\,,}\\
  \,{(-14.3\pm 0.8^{+6.7}_{-4.1})}\,{\rm GeV}^{-2} & \mbox{$J=\frac32$\,,}
	\end{cases}
\ee
where the first error is due to the statistical accuracy of the lattice
data in the region $r<1\,{\rm fm}$ and the second error is due to
the systematic uncertainty in the extrapolation for $r>1\,{\rm fm}$
{(with the uncertainties from Eq.~(\ref{Eq:integral-large-r}) 
combined in quadrature)}.

From Eqs.~(\ref{Eq:nu},~\ref{Eq:ratio-gs-gc},~\ref{Eq:integral})
we obtain the value for the chromoelectric polarizability
\be
	\alpha(1S) 
	=
	\begin{cases}
	\,({1.63 \pm 0.09{\,}^{+0.69}_{-0.44} \pm 0.44 \pm 0.11 \pm 0.01}
	)\,{\rm GeV}^{-3}
	& \mbox{$J=\frac12$\,,}\\ 
	\,({1.64 \pm 0.09{\,}^{+0.76}_{-0.47} \pm 0.44 \pm 0.11 \pm 0.01}
	)\,{\rm GeV}^{-3}
	& \mbox{$J=\frac32$\,,}
	\end{cases}	
\ee
with the errors due to the following uncertainties (in this order):
statistical accuracy of the lattice data in the region $r<1\,{\rm fm}$, 
systematic uncertainty of $\int\di^3r\,V_{\rm eff}(r)$ due to extrapolation
in the region $r > 1\,{\rm fm}$, uncertainty of the ratio $(g_c/g_s)^2$ 
and that of $\nu$, uncertainty of the lattice value for $M_N$
(the latter was not quoted in \cite{Sugiura:2017vks} but is estimated 
to be of the order of ${\cal O}(10\,{\rm MeV})$ \cite{Takuya}).
Combing the uncertainties in quadrature we obtain
\be
	\alpha(1S) 
	=
	\begin{cases}
	\,{(1.63 \pm 0.09{\,}^{+0.82}_{-0.63})}\,{\rm GeV}^{-3}
	& \mbox{$J=\frac12$\,,}\\ 
	\,{(1.64 \pm 0.09{\,}^{+0.89}_{-0.65} )}\,{\rm GeV}^{-3}
	& \mbox{$J=\frac32$\,.}
	\end{cases}
\ee
The agreement of the $\alpha(1S)$ values extracted from $V_{\rm eff}(r)$ 
in the $J=\frac12$ and $\frac32$ channels supports the assumption that 
heavy quark mass corrections do not play a dominant role in our analysis.
Rounding off and combing all sources (statistical and systematic) 
of uncertainties, we obtain for both channels
\be\label{eq:alpha-1S}
	\alpha(1S) = {(1.6 \pm 0.8)} \,{\rm GeV}^{-3} \;. 
\ee
We stress that this result has very little sensitivity to the shape of 
$V_{\rm eff}$ at small $r$ since we need the integral $\int\di^3r\,V_{\rm eff}(r)$ 
where the volume element suppresses the small-$r$ region. The result is 
much more sensitive to the large-$r$ dependence of $V_{\rm eff}$. We have
conservatively estimated the pertinent systematic uncertainty by assuming
extreme limiting cases in Eq.~(\ref{Eq:integral-large-r-prepare}). 
It is important to keep in mind that the result (\ref{eq:alpha-1S}) 
may have further systematic uncertainties inherent to the lattice data 
(discretization effects, unphysical light quark masses) which cannot be 
estimated at this point.

\section{Possibility for hadrocharmonia}

The charmonium-nucleon potential is attractive and we can study the 
possibility of a bound state -- hadrocharmonium \cite{Dubynskiy:2008mq}.
{A candidate for such a state with a mass around $4450\,{\rm MeV}$ 
was recently observed by LHCb \cite{Aaij:2015tga}.}
To do this we rescale the lattice effective potential by the factor 
$M_N^{\rm phys}/M_N^{\rm lattice}$, where $M_N^{\rm phys}=940$~MeV 
is physical nucleon mass and $M_N^{\rm lattice}=1816$~MeV the nucleon 
mass obtained in lattice measurements of \cite{Sugiura:2017vks} .
We need this rescaling to ensure the physical normalization 
condition (\ref{Eq:Veff-norm}) for the effective potential. 
 
Solving the Schr\"odinger equation for the rescaled potential we confirm 
the conclusion of Ref.~\cite{Sugiura:2017vks}  that $J/\psi$ does not
form the bound state with the nucleon. Now we can study the possibility 
of a nucleon bound state with $\psi(2S)$. To do this we note that 
according to Eq.~(\ref{Eq:Veff}) the shape of the nucleon-$\psi(2S)$ 
potential is the same as for the corresponding potential for $J/\psi$,
the only difference is the overall normalization factor due to 
chromoelectric polarizability.
 
Using the results for the shape of the effective potential extracted 
here from the lattice and treating $\alpha(2S)$ as a free parameter, 
we obtain the following results:
 \begin{itemize}
 
 \item 	The nucleon-$\psi(2S)$ bound states can form if $\alpha(2S)\ge
	\alpha_{\rm crit}(2S)={(8\pm4)}$~GeV$^{-3}$,
	where error bars are 
	due to statistical and systematic error of our fit, and due 
	to uncertainty of $(g_s/g_c)^2$, see Eq.~(\ref{Eq:ratio-gs-gc}).
	Note that in the ratio $\alpha(2S)/\alpha(1S)$ many systematic 
	uncertainties are canceled. For 
	this ratio we obtain $\alpha_{\rm crit}(2S)/\alpha(1S)=(5.0\pm0.5)$.  
	The values of $\alpha_{\rm crit}(2S)$ from the $J=1/2$ and $J=3/2$
	potentials are indistinguishable within error bars. 
 	The obtained value of $\alpha_{\rm crit}(2S)$ is compatible with 
	those obtained in Refs.~\cite{Eides:2015dtr,Perevalova:2016dln} 
	in completely different frameworks.  
	
 \item 	For $\alpha(2S)=({24\pm 12})$~GeV$^{-3}$ the bound state with mass 
	4450~MeV is formed. It may correspond to the narrow LHCb pentaquark
 	$P_c(4450)$. Again we have a good agreement with the findings of 
	Refs.~\cite{Eides:2015dtr,Perevalova:2016dln}.
	In terms of the ratio $\alpha(2S)/\alpha(1S)$ the hadrocharmonium 
	$P_c(4450)$ exists for $\alpha(2S)/\alpha(1S)=(15\pm1)$.
	Such a value of $\alpha(2S)$ and the results in
	Eqs.~(\ref{Eq:alpha-2S-to-1S-pheno},~\ref{eq:alpha-1S}) satisfy the 
	Schwarz inequality $\alpha(1S)\,\alpha(2S) \ge \alpha(2S\to1S)^2$
	\cite{Voloshin:2007dx}.
	
 \item 	From the  data \cite{Sugiura:2017vks} for $J=1/2$ and $J=3/2$ 
	effective potentials we are able to estimate the hyperfine splitting 
	between $\frac{3}{2}^-$ and $\frac{1}{2}^-$ components of  $P_c(4450)$. 
	We find the hyperfine mass splitting $(30\pm 30)\,{\rm MeV}$ 
	with tendency for $J=3/2$ to be heavier.
	This is compatible with both zero and with the estimate 
	of 5-10~MeV obtained in~\cite{Eides:2015dtr}.

\end{itemize}
We see that the lattice data of  \cite{Sugiura:2017vks} confirm 
the conclusions about nucleon-$\psi(2S)$ bound state made in 
Refs.~\cite{Eides:2015dtr,Perevalova:2016dln}.
It would be very interesting to make an independent lattice measurement 
of the  nucleon-$\psi(2S)$ effective potential.

\section{Conclusions}

The chromoelectric polarizability $\alpha(1S)$ of $J/\psi$ was
extracted on the basis of the formalism \cite{Voloshin:1979uv} 
from the lattice QCD data \cite{Sugiura:2017vks} 
on the effective nucleon-$J/\psi$ potential $V_{\rm eff}$. 
The final result is $\alpha(1S) = {(1.6 \pm 0.8)} \,{\rm GeV}^{-3}$.

The quoted error bar includes uncertainties due to 
strong coupling constants at nucleon and charmonium scales, 
parameter $\xi_s$ describing the fraction of baryon energy 
carried by gluons, and statistical error bars of the lattice 
data \cite{Sugiura:2017vks} on $V_{\rm eff}$ for $r\le 1\,{\rm fm}$.
In this region the systematic uncertainty due to choosing a specific 
fit Ansatz for $V_{\rm eff}$ is negligible because only the integral 
$\int\di^3r\,V_{\rm eff}(r)$ is needed for the extraction.
Exploring guidance from chiral symmetry (which dictates the behavior 
of $V_{\rm eff}$ at large $r$) we were able to provide a conservative
estimate of the systematic uncertainty due to extrapolation beyond 
$r > 1\,{\rm fm}$ where the lattice data for $V_{\rm eff}$ are 
compatible with zero or not available.

The extracted $\alpha(1S)$-value may have further systematic
uncertainties which cannot be estimated at this point, one of which
concerns our approach and the assumption of the heavy quark limit. 
The compatibility of lattice data for $V_{\rm eff}$ in the angular 
momentum channels $J=\frac12$ and $J=\frac32$ \cite{Sugiura:2017vks} 
provides an encouraging hint (but not more than that) that heavy quark 
mass corrections to $V_{\rm eff}$ might be within statistical error bars 
of the lattice data \cite{Sugiura:2017vks}. 
Unestimated potential systematic uncertainties pertain also to the
lattice data (discretization effects, unphysical light quark masses)
 \cite{Sugiura:2017vks}. 
Future lattice QCD studies will allow us to test whether the charm quark 
mass is large enough for the validity of our approach, and allow us to 
assess systematic uncertainties inherent to lattice simulations.

The obtained value $\alpha(1S) = {(1.6 \pm 0.8)} \,{\rm GeV}^{-3}$
is larger than the perturbative prediction
$\alpha(1S)_{\rm pert.} \approx 0.2\,{\rm GeV}^{-3}$  
\cite{Peskin:1979va,Eides:2015dtr}
which was so far basically the only available information on the
chromoelectric polarizability of $J/\psi$.
The larger value obtained here is in line with the suspicion 
$\alpha(1S)\gtrsim |\alpha(2S\to1S)|$ \cite{Sibirtsev:2005ex} 
with the value $|\alpha(2S\to 1S)| \approx 2\,{\rm GeV}^{-3}$ from 
$\psi^\prime\to J/\psi \, \pi\pi$ decays \cite{Voloshin:2007dx} (which may 
be reduced \cite{Guo:2006ya} by final state interaction effects).
This argument is not rigorous but based on the intuitive assumption that
off-diagonal matrix elements may be naturally expected to be smaller than
diagonal ones \cite{Sibirtsev:2005ex}.

We also studied the possibility of the nucleon-$\psi(2S)$ bound state.
We came to conclusions which are similar to those in 
Refs.~\cite{Eides:2015dtr,Perevalova:2016dln}, 
and support the interpretation of $P_c(4450)$ as $\psi(2S)$-nucleon 
bound state if $\alpha(2S)/\alpha(1S) \approx 15$. Our result is 
compatible with the value  of $\alpha(2S)\approx 17\,{\rm GeV}^{-3}$ 
obtained in Refs.~\cite{Eides:2015dtr,Perevalova:2016dln} 
in completely different frameworks.
This is remarkable, 
considering that in Refs.~\cite{Eides:2015dtr,Perevalova:2016dln} 
chiral models were used with massless \cite{Eides:2015dtr} and physical 
\cite{Perevalova:2016dln} pion masses, while here we used lattice data 
obtained at large unphysical $m_\pi$.
The results for the $\psi(2S)$ chromoelectric polarizability 
obtained in Refs.~\cite{Eides:2015dtr,Perevalova:2016dln} 
and here are based on the interpretation of $P_c(4450)$ as
a hadrocharmonium. Our analysis also provides
independent support for this interpretation.

The results obtained in this work contribute to a better understanding 
of the chromoelectric polarizabilities of charmonia, and will have
interesting applications for the phenomenology of hadrocharmonia.

\acknowledgments
We thank Takuya Sugiura for valuable correspondence 
and providing us with the data of Ref~\cite{Sugiura:2017vks}.
This work was supported in part by the National Science Foundation 
(Contract No.~1406298),
the Wilhelm Schuler Stiftung, 
and by CRC110 (DFG).

\newpage


\end{document}